\documentclass[prl,showpacs,twocolumn]{revtex4-1}
\usepackage{amsmath}
\usepackage{amscd}
\usepackage{wasysym}
\usepackage[ansinew]{inputenc}
\usepackage[T1]{fontenc}
\usepackage{ae,aecompl}
\usepackage{amsfonts}
\usepackage{amsthm}
\usepackage{dsfont}
\usepackage{graphicx}

\newcommand{\ud}{\mathrm{d}}
\newcommand{\ui}{\mathrm{i}}
\newcommand{\ue}{\mathrm{e}}

\newcommand{\rmi}{\mathrm{i}}

\newcommand{\R}{\mathds{R}}

\providecommand{\norm}[1]{\lVert#1\rVert}

\providecommand{\abs}[1]{\lvert#1\rvert}

\newcommand{\tr}{\operatorname{tr}}

\newcommand{\pa}{\partial}
\newcommand{\la}{\langle}
\newcommand{\ra}{\rangle}

\newcommand{\rf}[1]{(\ref{#1})}

\renewcommand{\Im}{\operatorname{Im}}
\renewcommand{\Re}{\operatorname{Re}}

\begin{document}
\title{Wave packet evolution in non-Hermitian quantum systems}

\author{Eva-Maria Graefe$^{1}$}
\author{Roman Schubert$^{2}$}
\address{${}^1$ Department of Mathematics, Imperial College London, London, SW7 2AZ, United Kingdom\\
${}^2$ Department of Mathematics, University of Bristol, Bristol, BS8 1TW, United Kingdom}
\begin{abstract}
The quantum evolution of the Wigner function for Gaussian wave packets generated by a non-Hermitian Hamiltonian is investigated. In the semiclassical limit $\hbar\to 0$ this yields the non-Hermitian analog of the Ehrenfest theorem for the dynamics of observable expectation values. The lack of Hermiticity reveals the importance of the complex structure on the classical phase space: The resulting equations of motion are coupled to an equation of motion for the phase space metric---a phenomenon having no analog in Hermitian theories. 
\end{abstract}
\maketitle
Effective non-Hermitian Hamiltonians have long been used for the description of a wide range of open quantum systems \cite{nherm}. Their applications range from chemical reactions to  ultra cold atoms and laser physics. Complex potentials for matter waves can be tailored experimentally using standing light waves \cite{Kell97}, and the complex Schr\"odinger equation appears in optics using materials with complex refractive index. The latter analogy was used in the recently reported first experimental realizations of PT-symmetric non-Hermitian Hamiltonians \cite{GuoRuet,Bend98}, which boosted the interest in the field further. While much attention has been paid to the theoretical study of example systems \cite{Nhermrec,nhbh}, the generic dynamical features remain hitherto mostly unexplored. The present paper aims to fill this gap by investigating wave packet dynamics for general non-Hermitian systems.

The time evolution of wave packets is a powerful tool for the understanding of both dynamical and stationary properties of Hermitian quantum systems \cite{WaveProp,Kram08}. Furthermore, it is a convenient way to investigate the semiclassical limit and thus forms the basis of many semiclassical methods. In what follows we shall generalize the fundamental ideas of wave packet dynamics to non-Hermitian systems. From this we derive classical equations of motion in the spirit of a generalized Ehrenfest theorem. This will pave the way for the development of a semiclassical framework for non-Hermitian quantum systems, or more generally, absorbing wave equations.

The semiclassical properties of non-Hermitian dynamics have recently been approached from various directions. Examples include the study of ray dynamics of absorbing wave equations for weak non-Hermiticities \cite{AbsWave}; the mean-field approximation for a non-Hermitian many-particle system \cite{nhbh}; the quantum classical correspondence for open quantum maps in the chaotic regime \cite{OpenMaps}; and complex extensions of the quantum probability distribution \cite{Bend10b}. In \cite{09nhclass} a coherent state approximation has been applied to non-Hermitian systems, and a generalized canonical structure involving a metric gradient flow has been identified. Here we go beyond this study by investigating general Gaussian states that are allowed to change their shapes during time evolution, which can be interpreted as a time-dependent metric on the corresponding classical phase space. To derive the classical evolution equations in the spirit of the Ehrenfest theorem, we study the quantum evolution equation for the Wigner function and take the semiclassical limit. This results in a new type of classical phase space dynamics in which the evolution equations for the phase space coordinates depend on the local metric, and vice versa. The complex structure of the classical phase space, which is rarely considered in Hermitian systems, thus becomes highly relevant in the presence of non-Hermiticity. The main dynamical effect of the anti-Hermitian part of the quantum Hamiltonian in this semiclassical approximation is a damping of the motion. This strengthens the often speculated connection to classical dissipation where non-Hermitian Hamiltonians are a recurrent theme in the search for quantum counterparts (see e.g., \cite{Raza} and references therein).

\emph{Gaussian coherent states.} 
For a general $n$-dimensional quantum system 
we study the evolution of initial wave packets of the form 
\begin{equation}
\psi(q)=\frac{(\det \Im B)^{1/4}}{(\pi\hbar)^{n/4}}\,\ue^{\frac{\ui}{\hbar}[P\cdot (q-Q)+\frac{1}{2} (q-Q)\cdot B(q-Q)]},
\label{eqn:psi}
\end{equation}
where $P,Q\in\R^n$, and $B$ is a complex symmetric matrix with positive definite imaginary part so that $\psi$ is localized around $q=Q$ and normalized to unity. 
he Gaussian states (\ref{eqn:psi}) with fixed $B$ form a submanifold of Hilberts space wich has a 
natural complex structure  associated with $B$. In the semiclassical limit this submanifold can be identified with the classical phase 
space, which thus inherits not only the symplectic  but also a metric structure. These relations become most transparent using a phase space representation \cite{Weyl}. In the following we will focus on the evolution of the Wigner function $W$ of 
$\psi$. The Wigner function alllows a direct computation of
expectation values via phase space integrals $\la \psi ,\hat A\psi\ra/\la\psi,\psi\ra=\la A\ra_W:=\int WA\ud z/\int W \ud z$, where $\hat A$ is the Weyl quantization of $A(z)$ and $z=(p,q)$ are canonical phase space coordinates. The Wigner function of a Gaussian state \rf{eqn:psi} is Gaussian: 
\begin{equation}
\label{eqn:wigner_gauss}
W(z)=(\pi\hbar)^{-n}\ue^{-\frac{1}{\hbar} (z-Z)\cdot G(z-Z)}.
\end{equation}
Here $Z=(P,Q)\in \R^n\times \R^n$ and the matrix $G$ is related to $B$ by  
\begin{equation}
\nonumber
G=\begin{pmatrix} I & 0 \\ -\Re B & I\end{pmatrix}\begin{pmatrix} (\Im B)^{-1} & 0 \\ 0 & \Im B\end{pmatrix}
\begin{pmatrix} I &-\Re B\\  0 & I\end{pmatrix}\,\, .
\end{equation}
The matrix $G$ is nondegenerate, positive, and symmetric, and thus acts as a \textit{metric} on phase space. 
It is also symplectic, i.e. it satisfies $G\Omega G=\Omega$, where  
\begin{equation}
 \Omega=\begin{pmatrix} 0 & -I\\ I & 0\end{pmatrix}
 \end{equation}
is the symplectic, or canonical, structure on phase
space. With such a metric we can associate a compatible complex structure $J$ with $J^2=-I$ and $ \Omega J= G$ \cite{CS}. The structure defined by $\Omega$, $G$, and $J$ turns the phase space into a K{\"a}hler manifold. 

The Wigner function \eqref{eqn:wigner_gauss} is localized of order 
$\sqrt \hbar$ around the maximum $Z=(P,Q)$. In the semiclassical limit $\hbar\to 0$
this collapses to a phase space point $Z$. Hence  
the expectation value of an observable $\hat A$ satisfies
\begin{equation}\label{eq:exp-Z}
\la \hat A\ra_{W}=A(Z)+O(\hbar)
\,\, 
\end{equation}
for $A(z)$  smooth. 
The metric $G$ describes the shape and orientation of $W$ in phase space, and  therefore determines the variance of observables: 
\begin{equation}\label{eq:variance-G}
(\Delta \hat A)^2_\psi=\frac{\hbar}{2}\nabla A(Z)\cdot G^{-1}\nabla A(Z)+O(\hbar^2)\,\, .
\end{equation}

In the Hermitian case an initially Gaussian state stays approximately Gaussian during the time evolution up to the Ehrenfest time \cite{WaveProp}. The center moves according to the classical canonical equations of motion $\dot Z=\Omega\nabla H$, and the evolution of the metric is governed by the linearized Hamiltonian flow around the classical trajectory. In the framework of the time-dependent variational principle \cite{TDVP} it has been shown that this dynamics can also be described by Hamiltonian equations of motion. Although it plays a central role in semiclassical methods involving families of coherent states \cite{TDVP,GM}, the metric is often little investigated, as it does not enter the dynamical equations for the phase space variables. We will see shortly that this is fundamentally changed in the presence of non-Hermiticity. 

\emph{Non-Hermitian Wigner-von Neumann equation.} 
Decomposing the Hamiltonian in its Hermitian and 
anti-Hermitian part $\hat H-\rmi\hat \Gamma$, where we assume $\hat H$ and $\hat \Gamma$ to be given as the Weyl quantizations
of sufficiently well-behaved classical observables $H(z)$ and $\Gamma(z)$, the evolution equation
for a density operator $\hat W$ follows from the Schr\"odinger equation as
\begin{equation}\label{eq:density-op-evolution}
\ui \hbar \pa_t \hat W=[\hat H,\hat W]-\ui [\hat \Gamma ,\hat W]_{+}\, ,
\end{equation}
where $[\cdot,\cdot]_{+}$ denotes the anti commutator. Thus, the evolution equation of a general Wigner function is given by
\begin{equation}
\label{eqn:wigner-evo}
\ui\hbar \pa_t W=(H\sharp W-W\sharp H)-\ui(\Gamma\sharp W+W\sharp\Gamma)\,\, , 
\end{equation}  
where $(A\sharp B)(z)$ denotes the Weyl product \cite{Weyl} for two phase space functions $A(z)$ and $B(z)$:
\begin{equation}
\nonumber
A\sharp B=A\ue^{\frac{\ui\hbar }{2} \overleftarrow{\nabla}_z\cdot \Omega \overrightarrow{\nabla}_{z}}B
\sim \sum_{k=0}^{\infty} \frac{1}{k!}\bigg(\frac{\ui\hbar }{2}\bigg)^k A\big(\overleftarrow{\nabla}_z\cdot \Omega \overrightarrow{\nabla}_{z}\big)^kB\,\, .
 \end{equation}
Here the arrows over the differential operators indicate whether they act on the function to the right or to the left.  We will now evaluate the leading order terms in $\hbar$ of \eqref{eqn:wigner-evo}. 
The Hermitian part of the evolution equation \rf{eqn:wigner-evo} is the well known Moyal bracket with an asymptotic expansion in odd powers of $\hbar$ whose leading term gives the Poisson bracket
\begin{equation}
H\sharp W-W\sharp H= \ui\hbar\nabla H\cdot \Omega\nabla W+O(\hbar^3)\,\, .
\end{equation} 
The anti-Hermitian part  has an expansion in even powers of $\hbar$, with the first two terms given by 
\begin{equation}
\Gamma\sharp W+W\sharp\Gamma  
=2\Gamma W -\frac{\hbar^2}{4}\Delta_{\Gamma}W+O(\hbar^4),
\end{equation}
where we introduced a second order differential operator defined by $\Gamma$ as $\Delta_{\Gamma}W:=\Gamma \big(\overleftarrow{\nabla}_z\cdot \Omega\overrightarrow{\nabla}_{z}\big)^2W$.
Denoting the matrix of second derivatives of $\Gamma(z)$ at $z$ by $\Gamma''(z)$ 
 the operator $\Delta_{\Gamma}$ can be written in the form
$\Delta_{\Gamma}=\nabla\cdot \Gamma_{\Omega}^{''}\nabla$, with $\Gamma_{\Omega}''(z):=\Omega^t\Gamma''(z)\Omega$.
It follows that $\Delta_{\Gamma}$ is Hermitian. Furthermore, 
if $\Gamma''(z)$ is symplectic, then
$\Gamma_{\Omega}''=\Gamma''^{-1}$, and $\Delta_{\Gamma}$ is the Laplace-Beltrami operator defined by $\Gamma''$. 

Summarizing, in leading order of $\hbar$ the dynamical equation 
for the Wigner function reads
\begin{equation}\label{eq:pphase-space-evo}
\hbar \pa_t W=-\bigg(-\frac{\hbar^2}{4}\Delta_{\Gamma}- \hbar \nabla H\cdot\Omega \nabla+2\Gamma \bigg)W\, . 
\end{equation}
For vanishing $\Gamma$ we recover the classical 
Liouville equation for the transport of phase space densities. 
For non-vanishing and positive $\Gamma$, on the other hand, the $\Gamma$ term defines a diffusion equation.

The higher order terms are
of order $\hbar^3\abs{\pa^3W}(\abs{\pa^3H}+\abs{\pa^3\Gamma})$, i.e. they are
small if the derivatives of $H(z)$, $\Gamma(z)$ and $W(z)$ do not grow too fast as $\hbar\to 0$. 
If the derivatives of $W$ are bounded, 
the first term on the right side of \eqref{eq:pphase-space-evo} is 
of lower order than the other terms. In this case the solution $W(t,z)$ is obtained 
by transporting the initial $W(z)$ along the Hamiltonian flow generated by $H$, multiplied by a damping factor, which is determined by the integral of $\Gamma$ along the Hamiltonian trajectories of $H$. This behavior is well known from 
damped wave equations. For a Gaussian initial state \eqref{eqn:wigner_gauss}, on the other hand, the term $\hbar^2\Delta_{\Gamma}W=O(\hbar)$ is of the same order as the Hamiltonian term in \eqref{eq:pphase-space-evo}, and the dynamics differ drastically from the Hermitian case. 

\emph{Gaussian evolution.}
In what follows we investigate the solution of \eqref{eq:pphase-space-evo} for an initial Gaussian Wigner function \eqref{eqn:wigner_gauss}. Inserting a Gaussian ansatz for the time evolved Wigner function 
\begin{equation}
\nonumber
W(t,z)=\frac{\alpha(t)}{(\hbar\pi)^n}\ue^{-\frac{1}{\hbar}\delta z \cdot G(t)\delta z}\,\, ,\,\, \text{with}\,\, \delta z:=z-Z(t)\,\,
\end{equation}
into \eqref{eq:pphase-space-evo} yields 
\begin{equation}\label{eq:ansatzIn}
\begin{split}
\bigg[\hbar\frac{\dot{\alpha}}{\alpha} &+2\dot{Z}\cdot G \delta z-\delta z\cdot \dot{G}\delta z\bigg]W(z)\\
=&\bigg[\delta z\cdot G\Gamma_{\Omega}''\delta z -2 \nabla H\cdot \Omega G\delta z\\
&\hspace*{1.5cm}-\frac{\hbar}{2}\tr\big(\Gamma_{\Omega}'' G\big)-2\Gamma\bigg]W(z)\,\, .
\end{split}
\end{equation}
Following the well established method of Heller and Hepp \cite{WaveProp},  
we expand $\Gamma(z)$ and $H(z)$ up to second order around
$z=Z$: $\Gamma(z)\approx\Gamma(Z)+\nabla\Gamma(Z)\cdot \delta z+\frac{1}{2}\delta z\cdot \Gamma''(Z)\delta z$ and $\nabla H(z)\approx \nabla H(Z)+ H''(Z)\delta z$. Since $W(z)$ is localized around $z=Z$ with a width proportional to
$\sqrt\hbar$ the remainder terms are of order $\hbar^{3/2}$.  
Separating different powers of $\delta z=z-Z$ in \eqref{eq:ansatzIn} then yields the
following three equations of motion for $Z(t)$, $G(t)$ and $\alpha(t)$:  
\begin{align}
\!\dot Z&=\Omega\nabla H(Z)-G^{-1}\nabla\Gamma(Z)\label{eq:theEq1}\\
\!\dot G&=H''(Z)\Omega G-G\Omega H''(Z)+\Gamma''(Z)-G\Gamma_{\Omega}''(Z)G\label{eq:theEq2}\\ 
 \!\frac{\dot\alpha}{\alpha} &=-\frac{2}{\hbar}\Gamma(Z)-\frac{1}{2}\tr\big[ \Gamma_{\Omega}''(Z)G\big]\,\, .\label{eq:theEq3}
\end{align}
To obtain \eqref{eq:theEq2} the symmetry enforcing convention $G=(G^t+G)/2$ was applied. As $W$ depends only on the symmetric part of $G$ any anti-symmetric part is unobservable. 

\begin{figure}[htb!]
\centering%
\includegraphics[width=4cm]{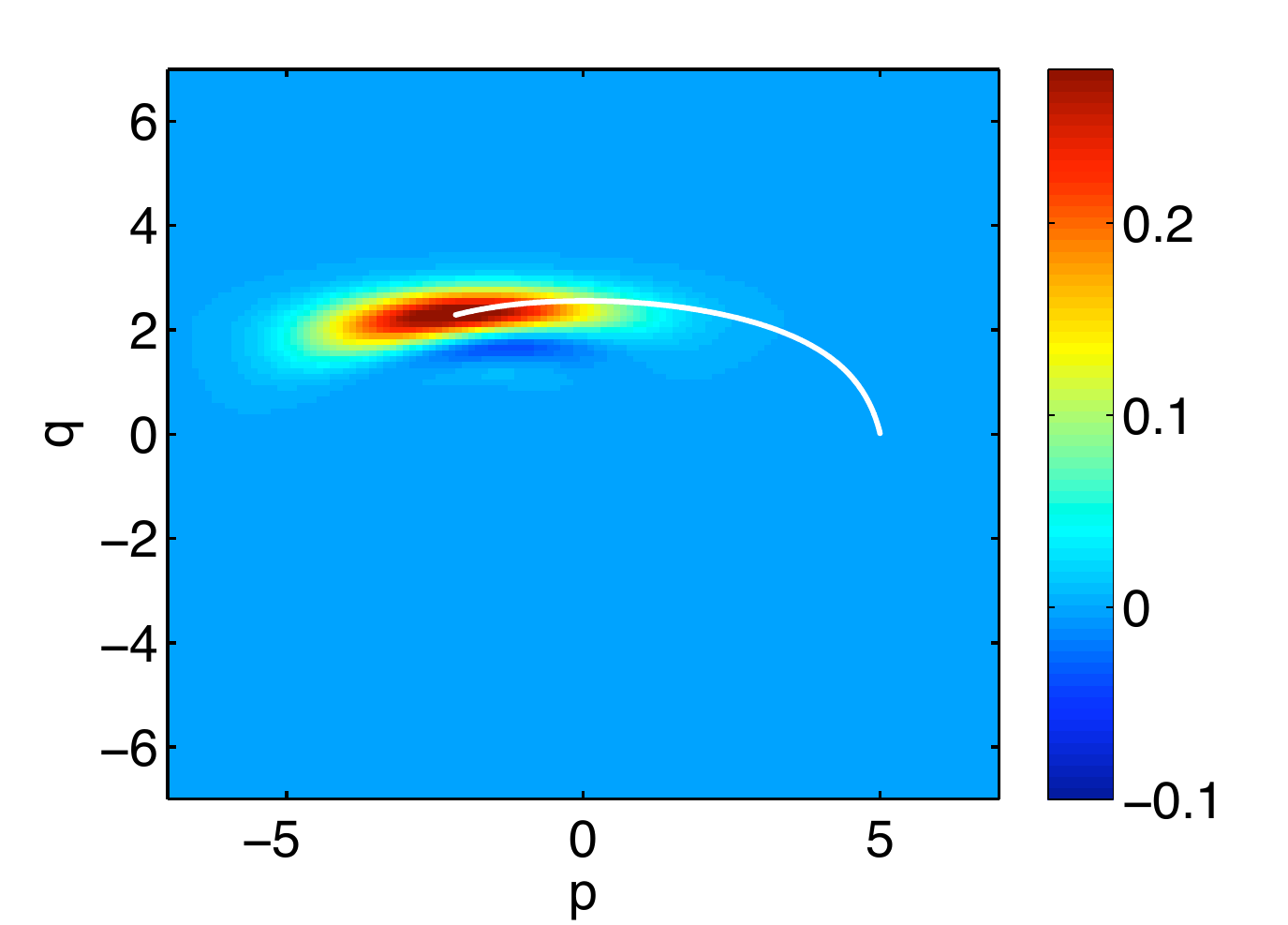}
\includegraphics[width=4cm]{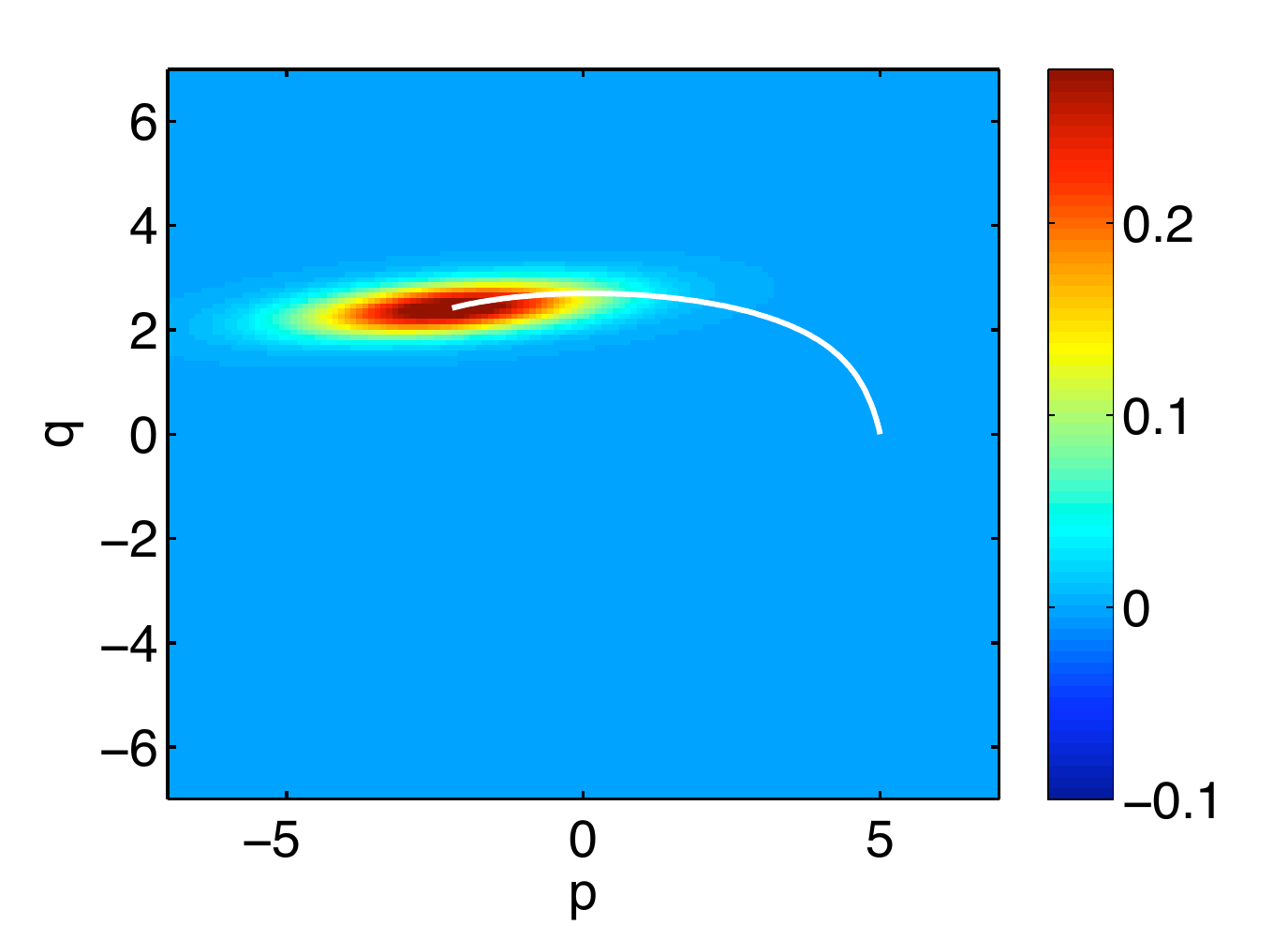}
\includegraphics[width=4cm]{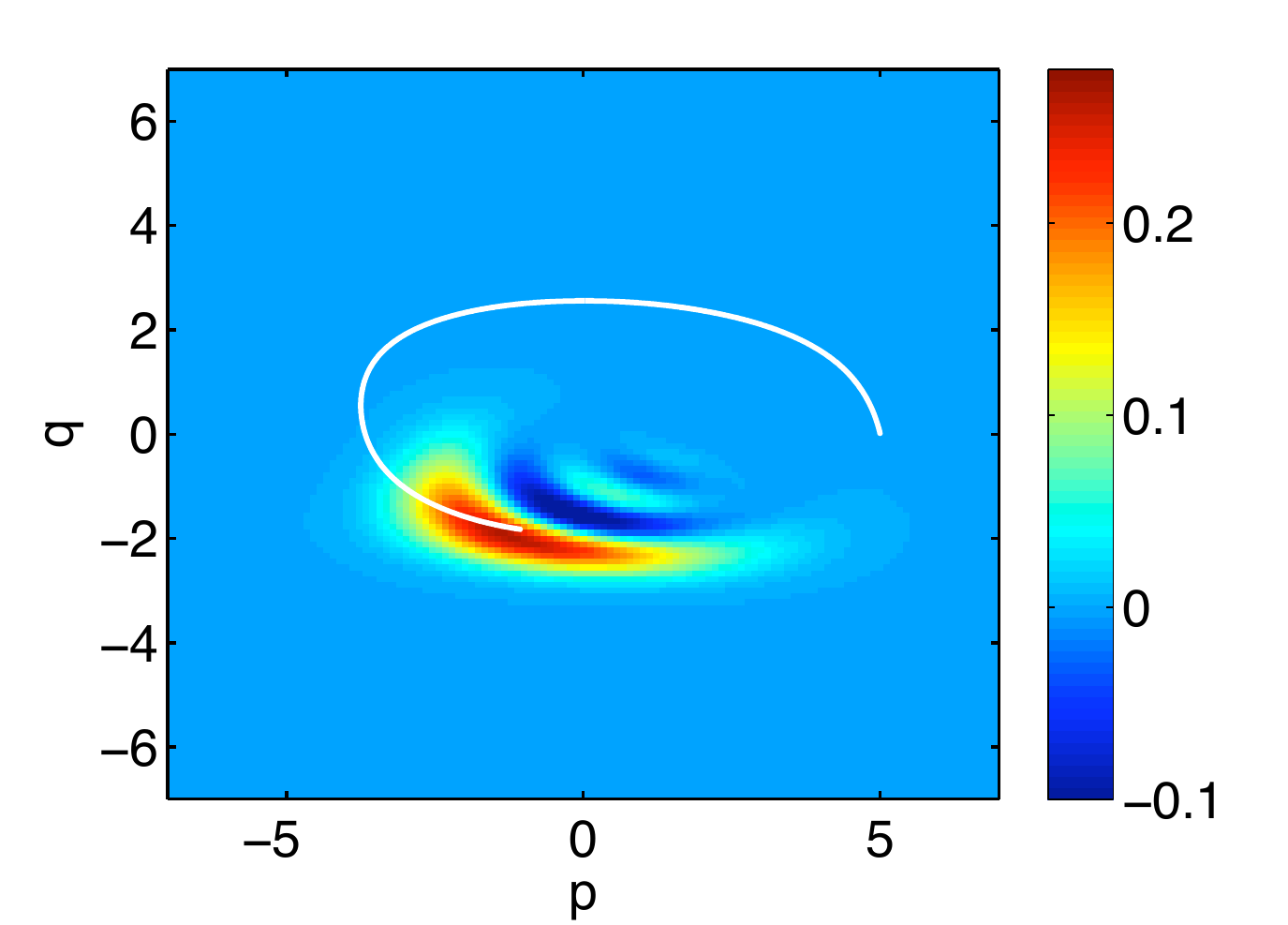}
\includegraphics[width=4cm]{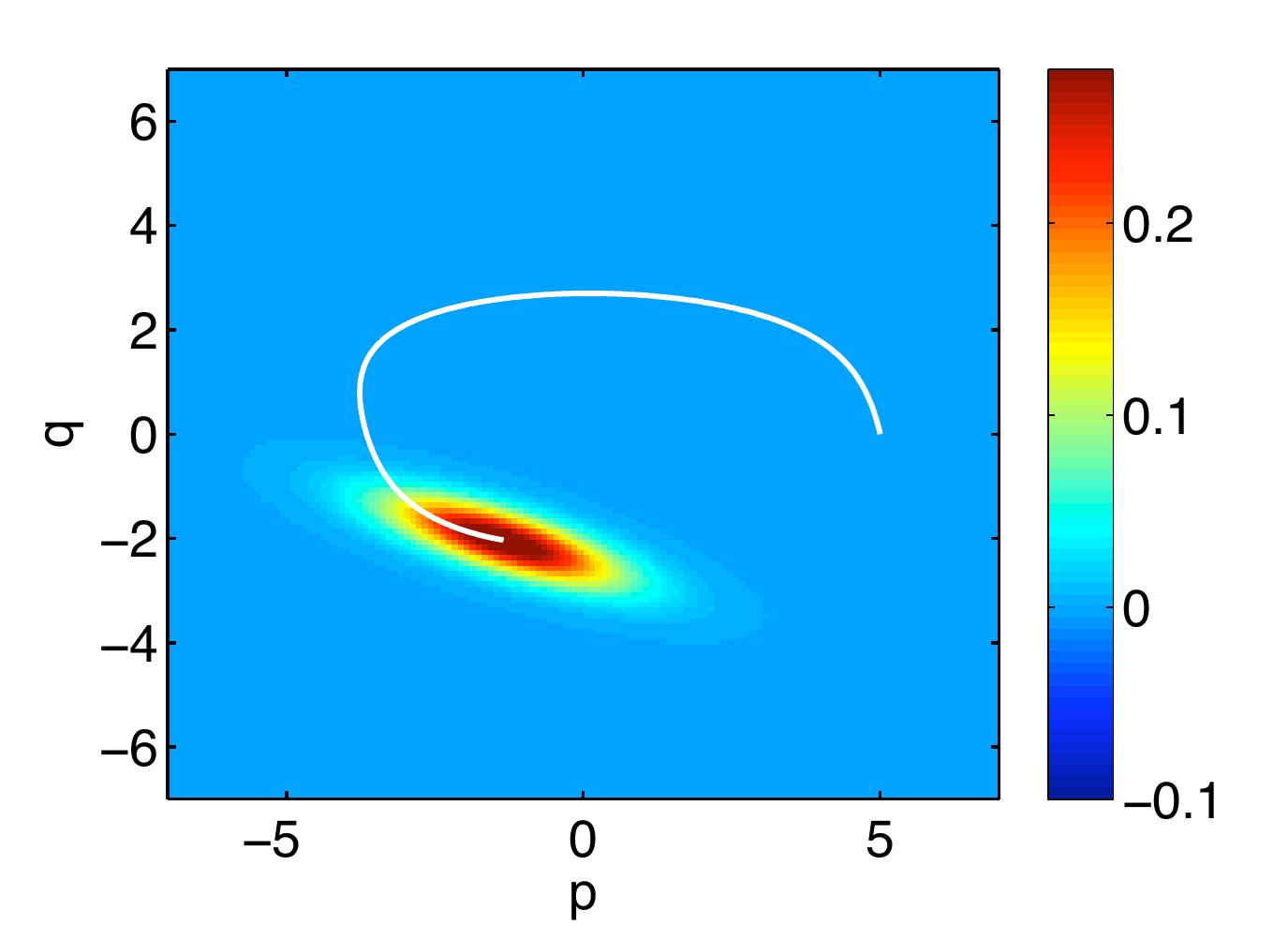}
\includegraphics[width=4cm]{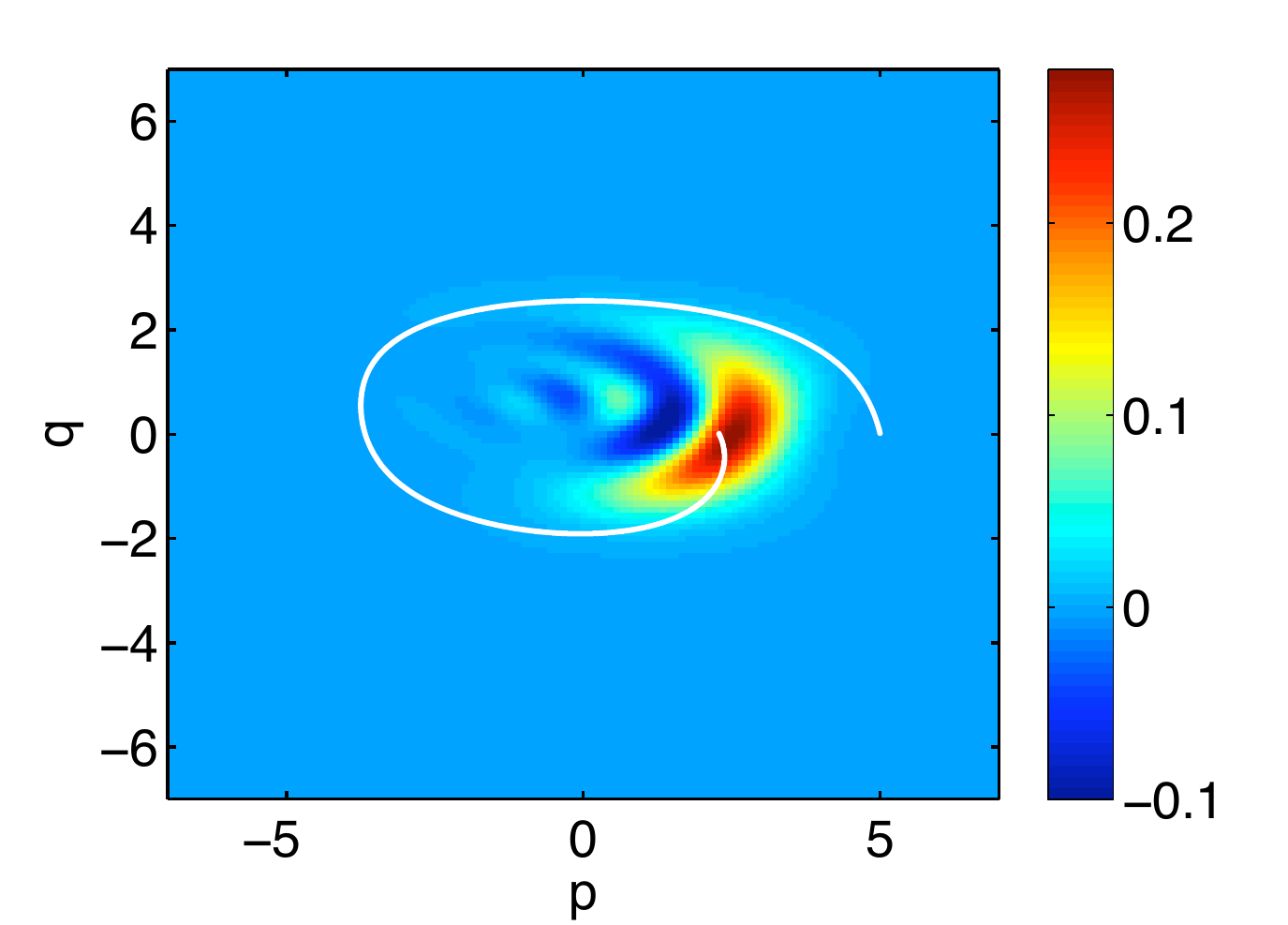}
\includegraphics[width=4cm]{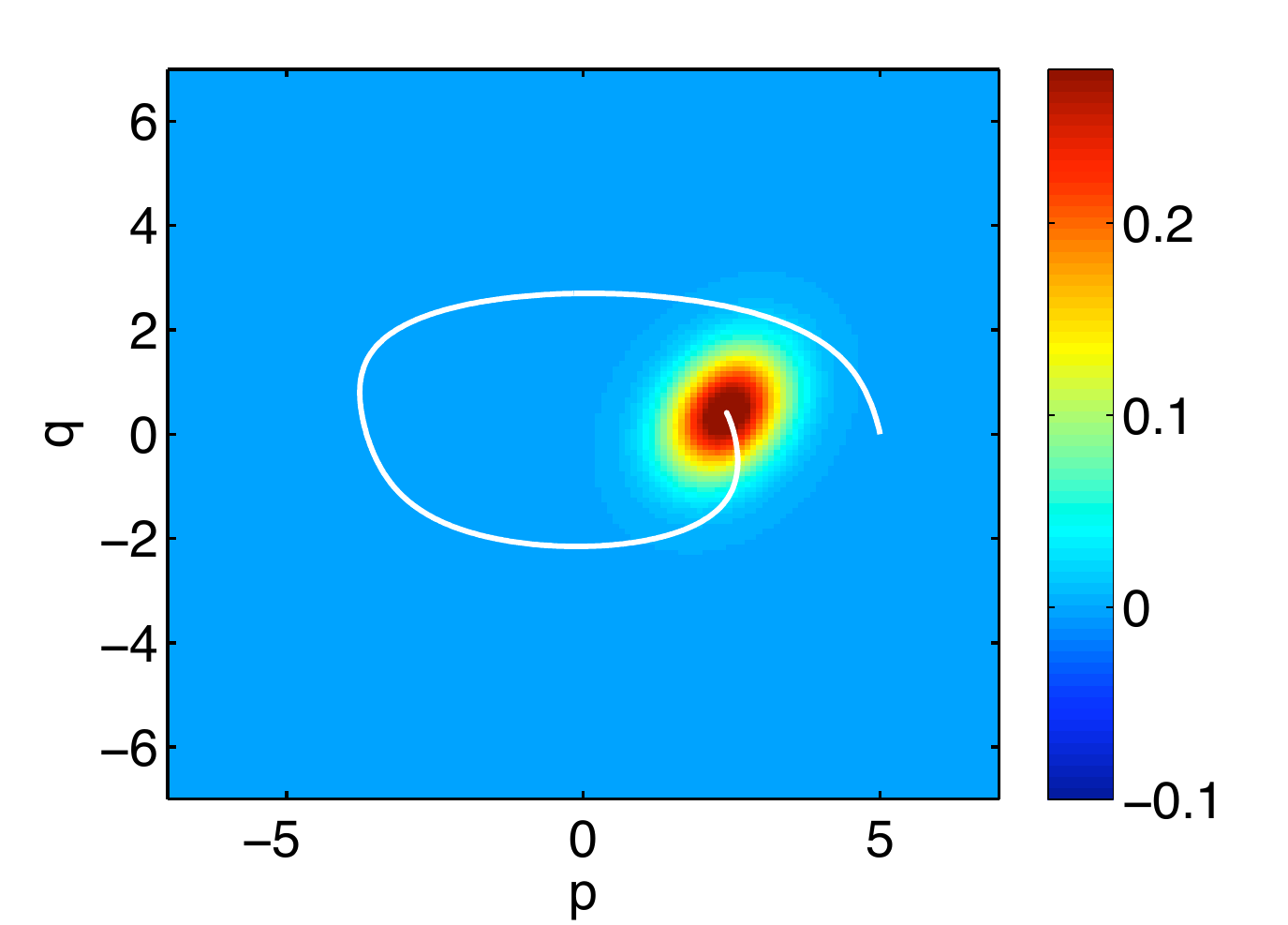}\\
\hspace{-0.35cm}
\includegraphics[width=3.5cm]{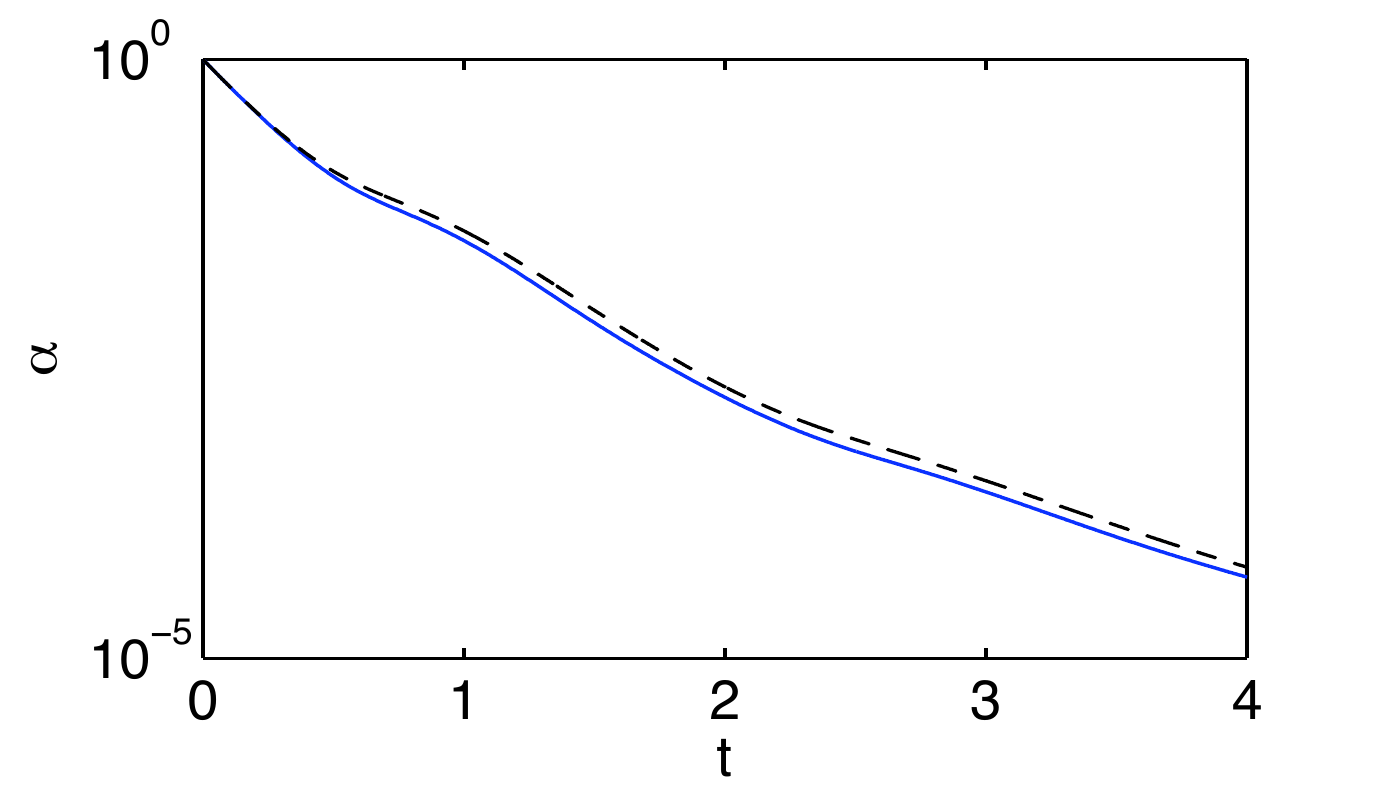}
\hspace{0.45cm}
\includegraphics[width=3.5cm]{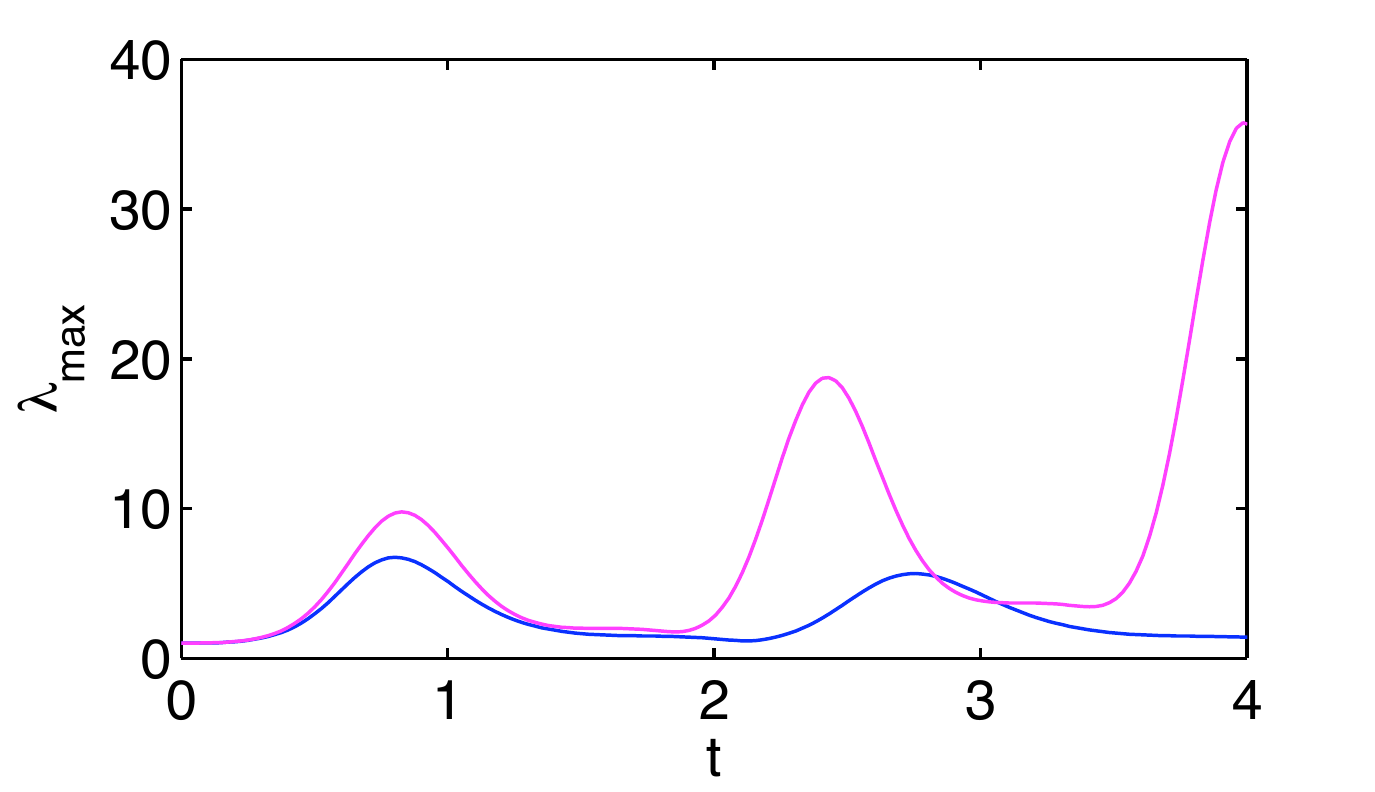}
\caption{Time evolution of the exact Wigner function (left column) and the semiclassical approximation (right column) for an initial state at $(p,q)=(5,0)$ at different times ($t= 1,\, 2.5,\, 4$) for the anharmonic oscillator. The white line shows the motion of the center. The left panel on the bottom shows the norm of the exact quantum state (black dashed line) and the semiclassical approximation (blue line), and the right panel shows the largest eigenvalue of $G(t)$ (blue line) in comparison with the Hermitian case $\gamma=0$ (pink line).}
\label{fig:Wignerqm3}
\end{figure}

The time evolution of expectation values and variances of arbitrary observables in Gaussian coherent states for small $\hbar$ is determined by $Z(t)$ and $G(t)$ according to \eqref{eq:exp-Z} and \eqref{eq:variance-G}. Equations \eqref{eq:theEq1}-\eqref{eq:theEq3} can be interpreted as the hitherto unidentified semiclassical limit of non-Hermitian quantum dynamics. This result goes beyond previous studies of the non-Hermitian Ehrenfest theorem \cite{Raza,MostaCurt,09nhclass} for two reasons. First, previous studies usually focussed on unnormalised expectation values, which prevented the identification of a classical structure, and second, disregarded the role of the metric, related to the widths of the quantum wave packet. The dynamics \eqref{eq:theEq1} emerging as the classical limit is no longer Hamiltonian, but has a Hamiltonian part  and a gradient part, determined by the Hermitian and anti-Hermitian 
parts of $\hat H-\ui\hat\Gamma$, respectively. The main dynamical effect of the anti-Hermitian part is to drive the motion towards the minima of $\Gamma$.  In addition, this gradient part is coupled to an evolution equation \eqref{eq:theEq2} for  the metric $G$  which in turn depends on \eqref{eq:theEq1}. In this context it is important to note that \eqref{eq:theEq2} preserves the symplectic nature of $G$ and 
hence describes an evolution of the complex structure on phase space. 
Further, the anti-Hermitian part leads to a change of the overall 
probability according to equation \eqref{eq:theEq3}, which can be interpreted as absorption or amplification. The first term gives the contribution from the center and the second term captures the influence of the width of the Wigner function. Note, however, that after renormalization the non-Hermitian Schr\"odinger equation is equivalent to norm-conserving nonlinear models for quantum dissipation \cite{Gisin}.

The quadratic approximation
around $z=Z(t)$ to $H(z)$ and $\Gamma(z)$ is expected to remain accurate so long as $W(t,z)$ stays strongly localized around $z=Z(t)$. Since for a symplectic $G$ we have $\norm{G^{-1}}=\norm{G}$, a suitable criterion for this is 
\begin{equation}\label{eq:T_Ehrenfest}
\hbar\norm{G(t)}\ll 1\,\, .
\end{equation}
The wave packet becomes delocalized at the Ehrenfest time $T_E$ defined by $\hbar \norm{G(T_E)}=1$ and the semiclassical approximation based on the central trajectory $Z(t)$ breaks down. 
The nonlinear term in the equation for $G(t)$ that is induced by $\Gamma$ can have a stabilizing effect on the long-time 
evolution of $G(t)$. Therefore, the non-Hermitian part can increase the Ehrenfest time, i.e. the time scale for which the semiclassical approximation is valid as compared to the Hermitian case.

\begin{figure}[htb!]
\centering%
\includegraphics[width=4cm]{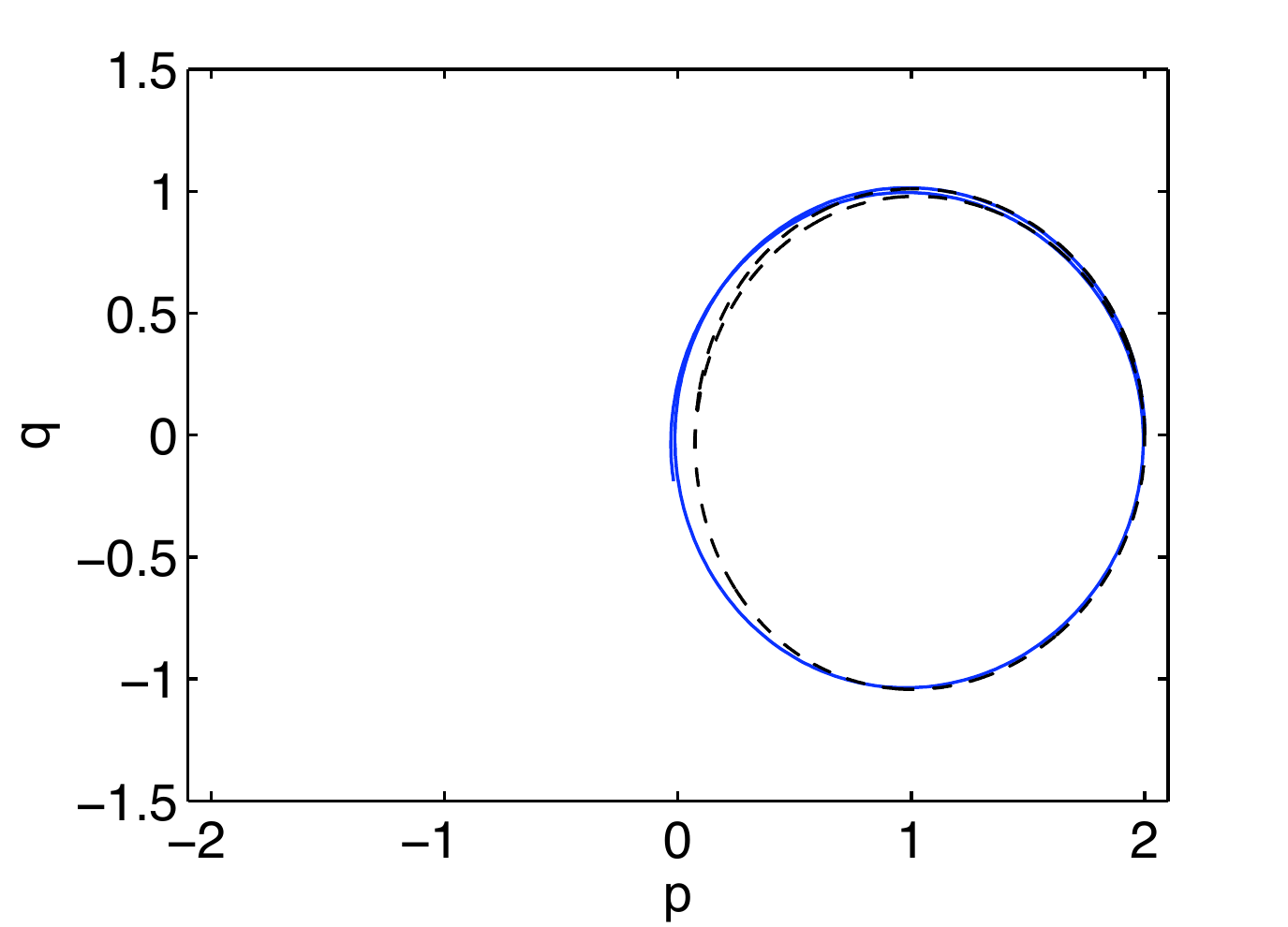}
\includegraphics[width=4cm]{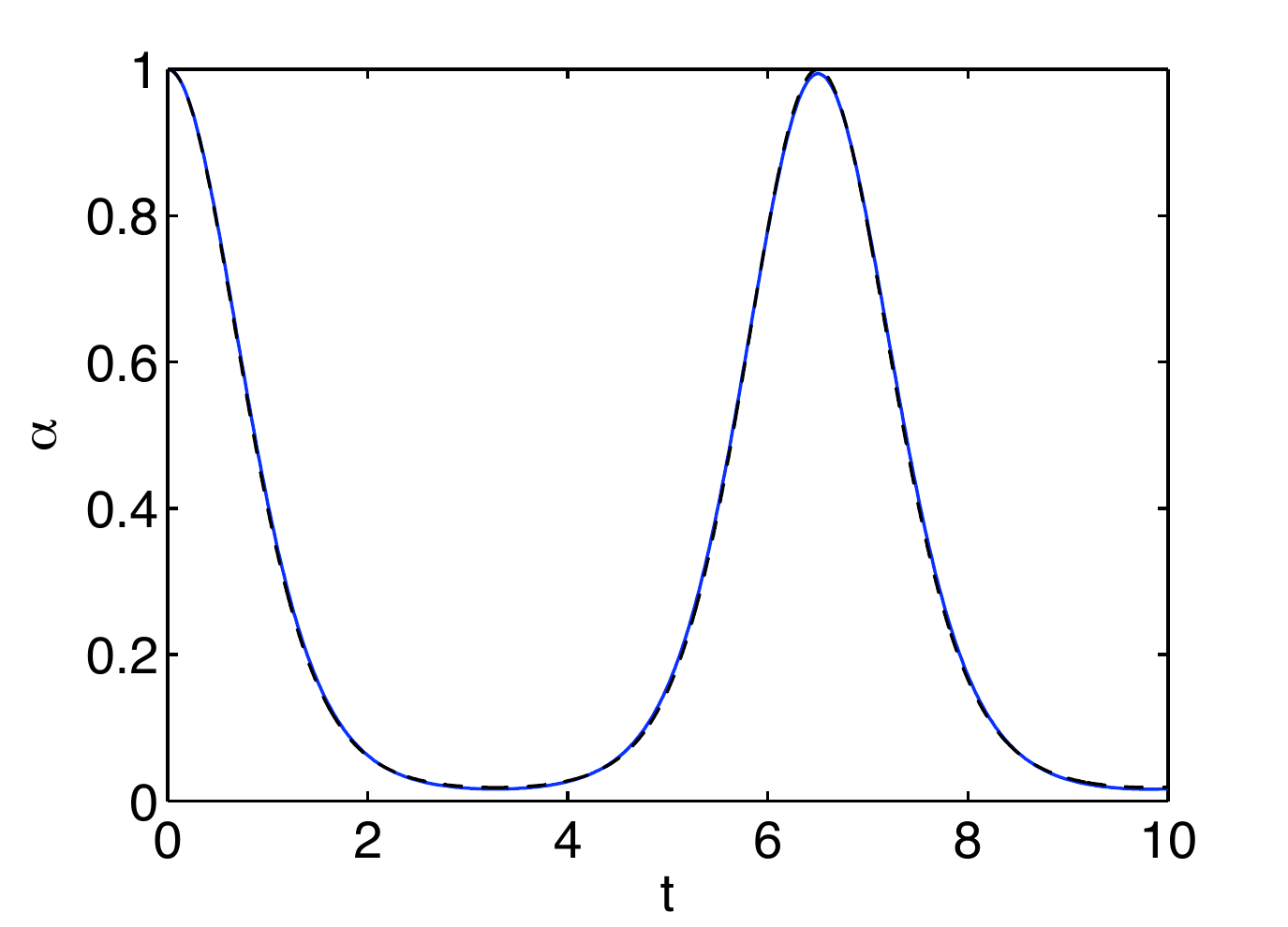}
\caption{Quantum evolution (black dashed line) versus semiclassical approximation (blue line) of a PT-symmetric waveguide for an 
initial state at $(p,q)=(0,2)$. Shown are the phase space evolution (left), and the evolution of the norm (right). } 
\label{fig:exp_q}
\end{figure}

\emph{Examples.} To illustrate our results we consider two examples. The first example is a non-Hermitian anharmonic oscillator with
$\hat H=\frac{\omega}{2}(\hat p^2+\hat q^2)+\frac{\beta}{4}\hat q^4$ and $\hat\Gamma=\frac{\gamma}{2}(\hat p^2+\hat q^2)$
respectively, with $\omega=1$, $\gamma=0.2$, and $\beta=0.5$. This can be interpreted as a quantum analogue of a 
damped anharmonic oscillator (see, e.g., \cite{Raza} and references therein). The simple structure of this model makes it an ideal 
testing ground for the semiclassical approximation proposed here. We set $\hbar=1$, which is equivalent to a rescaling upon 
which $\beta$ plays the role of an effective $\hbar$. For $\beta=0$ the semiclassical approximation becomes exact. While the 
Hermitian part tries to propagate a state along closed curves of constant energies around the origin, the anti-Hermitian part 
drives it towards the origin, thus acting as a damping. Figure \ref{fig:Wignerqm3} shows the exact numerical 
propagation and the semiclassical approximation for an initial Wigner function with $G_0=I$, which are in good agreement. Also the total mass of the exact state, a measure for the absorption, due to the anti-Hermitian part is well described by the semiclassical approximation $\alpha(t)$, as illustrated in the left panel on the bottom. The right panel shows the time-dependence of the larger eigenvalue 
of $G$ as a measure for $\norm{G(t)}$ in comparison with a Hermitian case $\gamma=0$. The result indicates that the Ehrenfest time, see \eqref{eq:T_Ehrenfest},  is increased in the non-Hermitian case, i.e. the semiclassical approximation is accurate over a longer time 
scale than in a comparable Hermitian case.

Second, we consider a simple model system for a PT-symmetric optical waveguide \cite{GuoRuet}: 
A single waveguide with harmonic confinement described by $H=\frac{1}{2}(\hat p^2+\hat q^2)$, and an anti-Hermitian part 
$\Gamma=5\tanh(0.2q)$ that models absorption on one side and equally strong amplification on the other side 
with a smooth transition in between. Although the Hamiltonian is complex, due to its special symmetry, it has real eigenvalues which can lead to a pseudo-closed behavior. This phenomenon is captured by our classical approximation. Figure \ref{fig:exp_q} shows an example of the full quantum evolution and its  classical counterpart. Both the phase space evolution and the dynamics of the norm are well approximated by the classical description. In particular, despite the anti-Hermitian part in the Hamiltonian, no sink of the dynamics is observed. Note that there is a stable fixed point at $(p,q)=(0,1)$, corresponding to the ground state of the quantum system.

\emph{Conclusion.}
The results presented here for the evolution of a Gaussian coherent state generated by a non-Hermitian Hamiltonian provide the basis for a more profound understanding of non-Hermitian time evolution, a topic of considerable interest in a wide range of subjects. In particular, the application to realistic examples of typically non local non-Hermitian Hamiltonians appearing in resonance physics is an interesting task for future studies. 
 
The authors thank D. C. Brody and H. J. Korsch for useful comments. EMG is supported by an Imperial College 
Junior Research Fellowship.

\end{document}